\documentstyle[12pt]{article}

\begin{document}
\rightline{FERMILAB-PUB-93/142-T}
\rightline{BNL-49447}
\bigskip\bigskip

\begin{center}
{\Large \bf Higgs Bosons at the Fermilab Tevatron} \\
\bigskip\bigskip

{\large\bf A.\ Stange and W.\ Marciano } \\
\medskip
Physics Department  \\
Brookhaven National Laboratory \\
Upton, NY\ \ 11973  \\
\bigskip\bigskip
and \\
\bigskip\bigskip
{\large\bf S.\ Willenbrock$^*$} \\
\medskip
Fermi National Accelerator Laboratory \\
P.~O.~Box 500 \\
Batavia, IL\ \ 60510 \\
\bigskip
and \\
\bigskip
Physics Department  \\
Brookhaven National Laboratory \\
Upton, NY\ \ 11973  \\
\end{center}
\bigskip\bigskip\bigskip
\footnotetext[1]{Address after September 1, 1993: Department of Physics,
University of Illinois, 1110 West Green Street, Urbana, Illinois 61801.}

\begin{abstract}
We study the production and detection of the standard-model Higgs boson at
the Fermilab Tevatron.  The most promising mode is $WH$ and $ZH$ associated
production followed by leptonic decay of the weak vector bosons and
$H\to b\bar b$.  It may be possible to detect a Higgs boson of mass
$m_H=60$--80 GeV with 1000 $pb^{-1}$ of integrated luminosity.
We also study the signature for a non-standard ``bosonic'' Higgs boson whose
dominant decay is to two photons.  A signal is easily established with
100 $pb^{-1}$ in the $WH$ and $ZH$ channels, with the weak vector bosons
decaying leptonically or hadronically, up to $m_H=100$ GeV.
\end{abstract}

\addtolength{\baselineskip}{9pt}

\section{Introduction}

The evidence is overwhelming that the electroweak interaction is
described by an SU(2)$_L\times{}$U(1)$_Y$ gauge theory, spontaneously
broken to U(1)$_{EM}$. However, the precise mechanism which breaks the
electroweak symmetry is unknown. The simplest mechanism is the standard
Higgs model, in which the symmetry is broken by the vacuum-expectation
value of a fundamental scalar field. This model predicts the existence
of a scalar particle, the Higgs boson, with fixed couplings to other
particles, but of unknown mass. The search for this particle constitutes
the baseline in our search for phenomena associated with the
electroweak-symmetry-breaking mechanism.

The current lower bound on the mass of the Higgs boson is about 60 GeV, from
the process $Z\to Z^*H$ at LEP \cite{LEP} ($Z^*$ denotes a virtual $Z$
boson). This bound is limited by statistics and backgrounds, and is
unlikely to improve
significantly. The next extension in reach will be provided by LEP II,
beginning in 1995, which will explore up to a Higgs-boson mass of about 80
GeV via $Z^*\to ZH$ \cite{LEP2}. Much higher masses will be explored by the
CERN
Large Hadron Collider (LHC) and the U.S. Superconducting Super Collider
(SSC), which can reach as high as $m_H=800$ GeV via the
processes $gg\to H$ and $V^*V^*\to H$ ($V=W,Z$) \cite{HHG}.

Conspicuously absent from this discussion is the Fermilab Tevatron $p\bar p$
collider.  It is generally assumed that the Tevatron cannot probe the
electroweak-symmetry-breaking mechanism.
However, there are two reasons why this is an appropriate time to study the
possibility of searching for the Higgs boson at the Tevatron. First, the
Tevatron is operating with high luminosity, and it now seems possible
that 100 $pb^{-1}$ of integrated luminosity will be delivered to each of the
two detectors by the end of 1994. Furthermore, with the Main Injector,
a five-fold increase in luminosity will be obtained, yielding 1000 $pb^{-1}$
of integrated luminosity per detector by the end of the century,
and perhaps more.
Second, it has been demonstrated that vertex detection is feasible in a
hadron-collider environment. This allows the detection of secondary
vertices from the decay of $b$ quarks, which is vital for searching for
the decay $H\to b\bar b$.

In Section~2 we study the production of the Higgs boson in association
with a $W$ or $Z$ boson at the Tevatron \cite{GNY}. The leptonic decay of
the $W$ or $Z$ boson provides a trigger for the event and suppresses the
backgrounds. In the mass range of interest, the standard Higgs boson decays
predominantly to $b\bar b$, and
vertex detection must be used to separate $b$ jets from light-quark
jets. This process has been extensively studied for the LHC \cite{LHC}
and SSC \cite{GKSG,SSC,BPP,K,AE}, but no similar study has been
undertaken for the Tevatron. While the analysis is similar to that of
the LHC/SSC in many respects, it differs in several important ways.

In Section~3 we discuss the search for a ``bosonic'' Higgs, which is a Higgs
boson with ordinary coupling to weak vector bosons, but suppressed coupling
to fermions. For $m_H<$ 100 GeV, this Higgs boson has a significant branching
ratio to two photons via a virtual $W$-boson loop.  Its production in
association with a $W$ or $Z$ boson yields an observable signal in both
the leptonic and hadronic decay modes of the weak vector bosons with
just 100 $pb^{-1}$ of integrated luminosity.  In Section~4 we summarize the
conclusions of our study.

\section{Standard Higgs boson}

The production of the Higgs boson at the LHC/SSC has been
extensively studied \cite{HHG}.
The same processes occur at the Tevatron, but with
different relative rates. We show in Fig.~1 the total cross sections for
various Higgs-boson production processes: gluon fusion via a virtual
top-quark loop \cite{GGMN}, associated production with a $W$ or $Z$ boson
\cite{GNY}, weak-vector-boson fusion \cite{CD}, and associated production
with a $b\bar b\;$ \cite{DW} or $t\bar t\;$ \cite{NZ,KUNSZT} pair
($m_t=150$ GeV).
We have included the QCD correction to the gluon-fusion cross section,
which is approximately a factor of 2.2 (in the $\overline{MS}$
scheme with $\mu=m_H$)
\cite{QCD,GSZ,DK}.\footnote{All
QCD corrections quoted in this paper are in the $\overline{MS}$
scheme, with a fixed set of next-to-leading-order parton distribution functions
\cite{HMRS} used for the leading-order and next-to-leading-order calculations.}
The QCD correction to $WH/ZH$ production is also
included; it is about $+25\%$ in the $\overline {MS}$ scheme with $\mu=M_{VH}$
\cite{HW}.
The QCD correction to vector-boson fusion is only a few percent \cite{HVW}.
Although gluon
fusion yields the largest cross section, it is relatively less important
than at the LHC/SSC due to the decreased gluon luminosity at the
Tevatron. The associated production with a $t\bar t$ pair is suppressed
relative to the LHC/SSC for the same reason, especially because the top
quark is relatively heavy compared with the machine energy. For this
reason, associated production with a $b\bar b$ pair is much larger than
with a $t\bar t$ pair, in contrast with the corresponding cross sections
at the LHC/SSC\null.

The branching ratios of the Higgs boson to $b\bar b$, $c\bar c$,
$\tau^+\tau^-$,
$WW^{(*)}$, and $ZZ^{(*)}$ \cite{R}
are shown in Fig.~2. The direct production of the Higgs boson via gluon
fusion, followed by $H\to b\bar b$ or $c\bar c$, is swamped by the
QCD production of $b\bar b$ and $c\bar c$ pairs.
Similarly, the mode $H\to WW^{(*)}\to \ell \bar\nu \bar
\ell^\prime \nu^\prime$ is swamped by the production of real
$W$-boson pairs. The $H\to ZZ^{(*)}\to \ell\bar\ell\ell^\prime
\bar\ell^\prime$ or $\ell\bar \ell\nu\bar \nu$ modes yield too few events
to observe.\footnote{This process was considered by V.~Barger and T.~Han in
Ref.~\cite{BH} at the Tevatron
with $\sqrt s = 3.6$ TeV and 1000 $pb^{-1}$, with a negative conclusion.}
The $H\to \tau^+\tau^-$ mode cannot be reconstructed due to the loss of the
tau neutrinos \cite{EHSV}.

Associated production with a $W$ or $Z$ boson is relatively more important
at the Tevatron than at the LHC/SSC.
The leptonic decay of the $W$ or $Z$ boson (including
$Z\to\nu\bar\nu$) provides a trigger for the event, and suppresses
backgrounds. It is these processes upon which we concentrate. The $WH$ process
has already been used at the Tevatron
to eliminate a very light Higgs boson \cite{CDF}. As we will show,
at least 1000 $pb^{-1}$ will be required to search for a more massive
Higgs boson at the Tevatron.
Thus we make cuts to simulate the acceptance of an upgraded CDF/D0
detector that one can envision existing by the time the Main Injector is
operating.

Consider the decay of the Higgs boson to a $b\bar b$ pair. To suppress
the $Wjj$ and $Zjj$ background, we must require that at least one
of the jets be identified as coming from a $b$ quark.
We require $|\eta_b|<2$ for the $b$ and $\bar b$
rapidity to simulate the coverage of the vertex detector (the coverage of the
existing CDF vertex detector is $|\eta_b|<1$), and we further require
the rapidity, transverse momentum ($p_T$), and isolation cuts
listed in Table 1.  The resulting cross sections are shown in Fig.~3,
including all branching ratios ($\ell=e,\mu$).  The $WH$ process
contributes to the $ZH$ cross section when the charged  lepton from the
$W$ decay is missed; the $ZH$ process contributes only a small amount
to the $WH$ cross section,
when one of the charged leptons from $Z\to  \ell\bar \ell$ is
missed.

\begin{table}[ht]
\begin{center}
\caption[fake]{Acceptance cuts used in section~2 to simulate an upgraded
CDF/D0 detector.}
\bigskip
\begin{tabular}{ll}
$|\eta_b|<2$ & $p_{Tb}>15$ GeV \\
$|\eta_\ell|<2.5$ & $p_{T\ell}>20$ GeV\\
${\not \!p}_{T}>20$ GeV &  (for $W\to\ell\bar\nu, Z\to\nu\bar\nu$) \\
$|\eta_j|<2.5$ & $p_{Tj}>15$ GeV \\
$|\Delta R_{b\bar b}|>0.7 $ & $|\Delta R_{b\ell}|>0.7$ \\
\end{tabular}
\end{center}
\end{table}

The principal backgrounds are $Wb\bar b$ \cite{KUNSZT,GKSG,AE,BSP,M}
and $Zb\bar b$ \cite{BSP,BM}, $WZ$ \cite{O} and $ZZ$ \cite{OO}
followed by $Z\to b\bar b$, and $t\bar t$ production. The $Wb\bar b$ and
$Zb\bar b$ backgrounds are shown in Fig.~3, assuming a $b\bar b$
invariant-mass resolution equal to a typical two-jet invariant-mass
resolution. We assume $\Delta E_j/E_j = .80/\sqrt E_j \oplus .05$
for the jet energy
resolution, which corresponds to approximately $\Delta M_{jj}/M_{jj}
= .80/\sqrt M_{jj} \oplus .03$
for the two-jet invariant-mass resolution.  We integrate the background
over an invariant-mass
bin of size $\pm2\Delta M_{jj}$ centered at $m_H$, which contains nearly
all the signal events.  The invariant-mass resolution may be degraded
somewhat since about forty percent of all events have at least one neutrino
from semileptonic $b$ decay. The signal-to-background ratio is of order unity,
and is better for $ZH$ than $WH$ because the $HZZ$ coupling is
bigger than the $HWW$ coupling by $M^2_Z/M^2_W$. Recall that at the
LHC/SSC the $ZH$ signal is swamped by $gg\to Zb\bar b$ \cite{GKSG}.
This is not the
case at the Tevatron; $q\bar q\to Zb\bar b$ is slightly larger than
$gg\to Zb\bar b$, due to the decreased gluon luminosity.

The background processes $WZ$ and $ZZ$, with $Z\to b\bar b$, populate the
region between about 80 and 100 GeV\null. The cross sections, including
the QCD corrections ($+33\%$ for $WZ$ \cite{O},
$+25\%$ for $ZZ$ \cite{OO} in the $\overline{MS}$
scheme with $\mu=M_{VV}$), are shown in
Fig.~3; they are comparable to the signal, and increase the
background in this region.  These backgrounds could be calibrated using
the purely leptonic decays of the gauge-boson pairs.

The top quark is also a potential background. The process $t\bar t\to
W^+W^-b\bar b$ mimics the $WH$ signal if one $W$ goes undetected.  We
assume a coverage for jets of $|\eta_j|<2.5$ and $p_{Tj}>15$ GeV,
and reject events with additional jets.\footnote{Rejecting
events with additional jets decreases the signal
somewhat.  This reduction can be minimized by increasing the minimum
$p_T$ of the additional jets, without greatly increasing the background.}
With this coverage, the dominant $t\bar t$ background occurs when the
charged lepton from a $W$ decay goes outside the coverage of the
detector.  We make the additional requirement that the transverse mass of the
observed charged lepton plus the
missing $p_T$ be less than $M_W$,
which is always true for the signal.  This cut reduces the background by about
a factor of two.  This background is shown
in Fig.~3(a) for $m_t = 150$ GeV; it
is far less troublesome than the direct $Wb\bar b$ background. The top quark
is also a background to the $ZH$ signal, with $Z\to \nu\bar \nu$, if both
$W$'s are missed.  This background is shown in Fig.~3(b); it is negligible.
In both cases, the top-quark
background decreases for a heavier top quark.  The top-quark background is
a much worse problem at the LHC/SSC.\footnote{The top-quark background at
the SSC was considered by R.~Kauffman in Ref.~\cite{K}.}

Another potential background is $Wc\bar c$ and $Zc\bar c$. Since the
charm-quark lifetime is comparable to that of the bottom quark, these
events can also produce a displaced vertex, and could as much as
double the background.  Final states with a single charm quark, such
as $W^-c\bar s$, can also contribute to the background if only one jet is
required to have a displaced vertex.
In Ref.~\cite{SSC,BPP} the charm-quark background was suppressed by demanding
that the $b$ quark decay semileptonically, with lepton momentum
transverse to the $b$ jet of at least 1 GeV\null. Due to the modest
number of signal events, one may not be able to afford such a cut at the
Tevatron.

One must also consider the $Wjj$ and $Zjj$ backgrounds, where the jets
come from light quarks \cite{EKS,GKS,EG,AR,BHOZ,BGKKS,MP2}.
Applying the same acceptance cuts and invariant-mass
resolution as before, we find that these cross sections are over one hundred
times as large as the $Wb\bar b$ and $Zb\bar b$ backgrounds.\footnote{The
$Wjj$ and $Zjj$ backgrounds were calculated using the code developed in
Ref.~\cite{BHOZ}.}
Excellent light-quark-jet/$b$-jet discrimination
will be required to eliminate this background.
The $Wjj$ and $Zjj$  backgrounds are much more
severe at the LHC/SSC because of the large gluon luminosity.  They were
eliminated in Ref.~\cite{SSC,BPP} by demanding a semileptonic decay of a
$b$ quark, with lepton momentum
transverse to the $b$ jet of at least 1 GeV, as mentioned above.

The efficiency for detecting a displaced vertex from a $b$-quark jet within
the coverage of the vertex detector and with $p_{Tb}>15$ GeV
is hoped to reach about 30 percent, or 50 percent to detect at least one
displaced vertex per $b\bar b$ pair.
We present in Table 2 the number of signal and background
events with at least one displaced vertex for various values of the
Higgs-boson mass for 1000 $pb^{-1}$ of integrated luminosity.  We assume a one
percent misidentification of a light-quark (or gluon) jet as a $b$ jet.
A charm jet produces a displaced vertex which mimics a $b$ jet with only five
percent probability, so the
$Wc\bar c$ and $Zc\bar c$ backgrounds are small compared with $Wb\bar b$ and
$Zb\bar b$.
The $Wjj$ and $Zjj$ backgrounds,
as well as the single-charm background (which we estimate is small),
could be eliminated completely by
requiring a double $b$ tag; however, the double-tag efficiency is only 10
percent per $b$ jet. The significance of the signal is a bit better with
a double tag, but the number of signal events is small.  Additionally, one
can tag $b$ jets using semileptonic decays; however, this has an
efficiency of only about 10 percent, with a one percent misidentification of
light-quark jets.

The number of signal events with at least one displaced vertex
for $m_H=60$--80 GeV may be enough to detect
the Higgs boson at the Tevatron, especially if the light-quark-jet
misidentification can be reduced below one percent.  The statistical
significance of the $WH/ZH$ signal is about $2.5/3.3\sigma$ for
$m_H=60$ GeV, and about $1.6/2.2\sigma$ for $m_H=80$ GeV.  If the
light-quark-jet
misidentification can be reduced to one-half percent, the significance
of the $WH/ZH$ signal increases to about $3.3/4.3\sigma$ and $2.1/2.9\sigma$
for $m_H=60$ and 80 GeV, respectively.  The significance of the $ZH$ signal is
greater than that of the $WH$ signal because the $HZZ$ coupling is bigger
than the $HWW$ coupling by $M_Z^2/M_W^2$, as we remarked earlier.
One should keep in mind
that LEP II
will have already explored these masses by the time 1000 $pb^{-1}$ is
delivered.  However, the confirmation of the Higgs boson at the Tevatron
would be of interest since both the $ZH$ (similar to LEP II) and $WH$
processes could potentially be detected.  A Higgs boson of mass near the
$Z$-boson mass will be difficult to separate from the $WZ$ and $ZZ$
backgrounds. A Higgs boson of mass in excess of the $Z$-boson mass would
require increased integrated luminosity for discovery.

\begin{table}[hbt]
\caption[fake]{Number of signal and background events, per 1000 $pb^{-1}$,
for production of the Higgs boson in association with a weak vector boson,
followed by $H\to b\bar b$ and $W\to \ell\bar \nu$,
$Z\to \ell\bar \ell,\nu\bar \nu$.}
\bigskip
\begin{center}
\begin{tabular}{cccc}
$m_H$ (GeV) & $WH/ZH$ & $Wb\bar b/Zb\bar b$ & $Wjj/Zjj$ \\
60 & 50/58 & 60/52 & 340/260  \\
80 & 28/35 & 39/38 & 260/210  \\
100 & 15/20 & 25/26 & 180/160 \\
120 & 8/10 & 17/18  & 130/120 \\
140 & 3/3.8 & 11/13 & 90/88 \\
\end{tabular}
\end{center}
\end{table}

Let us also consider the Higgs-boson signal for 100 $pb^{-1}$ of
integrated luminosity.  Presently only CDF has a vertex detector, so we
apply cuts to  simulate their acceptance: $|\eta_b|<1$, $p_{Tb}>15$ GeV,
$|\eta_e|<2.5$, $|\eta_\mu|<1$, $p_{T\ell}>20$ GeV,
${\not \!p}_{T}>20$ GeV, $|\Delta R_{b\bar b}| > 0.7$, $|\Delta R_{b\ell}|
> 0.7$.
Although the signal-to-background ratio is
again of order unity (for the  $Wb\bar b$ and $Zb\bar b$ backgrounds),
the number of signal events is only 2/2 for $WH/ZH$ (including the 50
percent single-tag efficiency) for $m_H= 60$ GeV, decreasing to 1/1
for $m_H=80$ GeV, not enough for discovery.

For $m_H>140$ GeV, the $H\to WW^{(*)}$ branching ratio becomes significant,
and one can consider searching for $WH$ production, followed by leptonic
decay of the like-sign $W$ bosons, leading to an isolated like-sign-dilepton
plus missing $p_T$ signature.  This signal has no irreducible background,
and could potentially be used for $m_H=140$--180 GeV.  Unfortunately,
the number of events in 1000 $pb^{-1}$ of integrated luminosity is
of the order of unity.  This signal could be useful for larger
integrated luminosity.

The search for the Higgs boson at the Tevatron is challenging, but is
matched by the importance of the search. A Higgs boson in the range
80--140 GeV (the so-called intermediate-mass range) is also difficult
for the LHC/SSC to discover. Studies at the LHC/SSC usually
focus on the rare two-photon decay of the Higgs boson \cite{AACHEN,TDR}.
It might be worthwhile
to reconsider the $b\bar b$ decay mode at these machines \cite{GKW,DGV}.

\section{``Bosonic'' Higgs}

The standard-model Higgs boson is responsible for generating the masses of
both the weak vector bosons and the fermions. One can imagine that the
mass generation of the weak vector bosons has little or nothing to do
with that of the fermions \cite{HKS,PWY,BC,BDHR,BBDKW,GVW,BK}. A Higgs boson
associated only with the generation of the weak-vector-boson masses would be
expected to have couplings to the weak vector bosons of standard-model
strength, but suppressed couplings to fermions. We will refer to such a
particle as a ``bosonic'' Higgs.\footnote{Such a particle is referred to
as a ``fermiophobic'' Higgs in Ref.~\cite{PWY}.}
For example, a bosonic Higgs can arise
in models with two Higgs doublets \cite{HKS,PWY,BDHR,BBDKW}
or with doublets and triplets\footnote{The
bosonic Higgs bosons in this model are the $H_5$ and the $H_1^{0\prime}$.}
\cite{GM,GVW,BK}, although
fine tuning of the renormalized coupling of the Higgs boson to fermions
is necessary in both cases \cite{HKS,BBDKW,GVW,BK}.  In general, one expects
some mixing to occur such that a bosonic Higgs has a non-vanishing coupling
to fermions.

Since the fermionic decay modes of a bosonic Higgs are greatly
suppressed, the decay of a bosonic Higgs of mass less than $2M_W$ is not
dominated by $H\to b \bar b$. It was noted in Refs.~\cite{PWY,BDHR,BK} that the
dominant decay mode of a sufficiently light bosonic Higgs
is to two photons via a
$W$-boson loop \cite{VVZS}, as shown in Fig.~4(a). The bosonic Higgs can
also decay to $b\bar b$ at one loop, as shown in Fig.~4(b); however, this
decay mode is suppressed relative to the two-photon mode by
$m^2_b/M^2_W$ (assuming the fine tuning mentioned above),
and can be neglected. As the Higgs-boson mass approaches
$M_W$, the decay $H\to WW^*$ (and even $H\to W^*W^*$ \cite{BBHK}) begins
to compete with the two-photon decay. The branching ratios of a bosonic
Higgs to $\gamma\gamma$, $WW^{(*)}$, and $ZZ^{(*)}$ are shown in Fig.~5. Note
that a top-quark loop does not contribute to the $\gamma\gamma$ decay of
the bosonic Higgs.  Charged Higgs bosons (present in multi-Higgs models) may
contribute, if they are not too heavy; their contribution is suppressed
relative to that of the $W$-boson by $M_W^2/m_{H^{\pm}}^2$.

A bosonic Higgs of mass less than about 60 GeV would have been seen at LEP
via $Z\to Z^*H$, with $H\to\gamma\gamma$ \cite{LEP3}. We will show that the
Tevatron, with 100 $pb^{-1}$ of integrated luminosity, can search for
the two-photon decay of the bosonic Higgs up to about $m_H=100$ GeV,
prior to the commissioning of LEP II.
The production process is the same as in the previous section: $q\bar
q\to WH,ZH$. The photons from the Higgs decay can serve as the
trigger, so we can consider both the leptonic and hadronic decays of the
$W$ and $Z$ bosons. We simulate the acceptance of the detector with the
cuts listed in Table 3.
Decreasing the rapidity coverage of all particles to
1 unit decreases the cross sections by about a factor of two. We show in
Fig.~6 the cross sections for $WH$ and $ZH$, followed by
$H\to\gamma\gamma$, for both (a) leptonic (including $Z\to \nu\bar\nu$)
and (b) hadronic decays of the $W$ and $Z$.
For the hadronic decays, we combine the
$WH$ and $ZH$ signals since the assumed two-jet invariant-mass resolution
(discussed below) cannot separate the $W$ and $Z$ peaks.  We use a machine
energy of $\sqrt s = 1.8$ TeV throughout this section.

\begin{table}[h]
\caption[fake]{Acceptance cuts used in section 3.}
\bigskip
\begin{center}
\begin{tabular}{ll}
$|\eta_\gamma|<2.5$ & $p_{T\gamma}>10$ GeV \\
$|\eta_\ell|<2.5$ & $p_{T\ell}>20$ GeV \\
${\not \!p}_T>20$ GeV & (for $W\to\ell\bar\nu, Z\to \nu\bar\nu$) \\
$|\eta_j|<2.5$ & $p_{Tj}>15$ GeV \\
$|\Delta R_{jj}|>0.7$ & $|\Delta R_{j\gamma}|>0.7$ \\
\end{tabular}
\end{center}
\end{table}

For the leptonic decay of the $W$ and $Z$ bosons, the main background is
$W\gamma\gamma$ \cite{KKS} and $Z\gamma\gamma$. We assume a photon energy
resolution of $\Delta E_\gamma/E_\gamma=.15/\sqrt{E_\gamma} \oplus .01$;
this corresponds to
a two-photon invariant-mass resolution of approximately $\Delta
M_{\gamma\gamma}/M_{\gamma\gamma}=.15/\sqrt{M_{\gamma\gamma}} \oplus.007$.
We integrate the background over an
invariant-mass bin of $\pm2\Delta M_{\gamma\gamma}$ centered about $m_H$,
which contains
nearly all the signal events. The resulting backgrounds are less than $10^{-4}$
$pb$ for $m_H>60$ GeV, too small to display in Fig.~6.  There is no background
from $WZ$ or $ZZ$ since $Z\to \gamma\gamma$ is forbidden by Yang's theorem
\cite{Y}.
Thus the signal for the bosonic Higgs produced in
association with a $W$ or $Z$ which decays leptonically is essentially
background free.

The dominant background when the $W$ and $Z$ decay hadronically is a
mixed QCD/QED process leading to a $jj\gamma\gamma$ final state
\cite{CALKUL,GK,MP,BHZO}. We take
a two-jet invariant-mass resolution as in the previous section, $\Delta
M_{jj}/M_{jj}=.80/\sqrt{M_{jj}} \oplus .03$, and integrate over a bin of width
$\pm2\Delta M_{jj}$ centered on the $W$ or $Z$ mass;
effectively, this corresponds to 65 GeV$<M_{jj}<$ 105 GeV,
with no effort made to separate the $W$ and $Z$ peaks.
The two-photon invariant-mass
bin is treated as above. The resulting cross section is shown in Fig.~6(b).
The signal is above the background up to a Higgs-boson mass of about
110 GeV.

Although the bosonic Higgs cannot be produced via gluon fusion, it is
produced with standard-model strength via $V^*V^*\to H$ ($V=W,Z$).
Because the virtual vector bosons are radiated from quarks and
antiquarks, the final state contains two jets, and thus also has a
$jj\gamma\gamma$ signal. This cross section is shown in
Fig.~7, with the acceptance cuts listed in Table 3,
plus the requirement that the two-jet
invariant mass exceed 100 GeV to separate it from the $WH/ZH$ signal.  Also
shown in Fig.~7 is the $jj\gamma\gamma$ background with the same cuts; the
signal lies above the background up to $m_H=90$ GeV.
The two-photon signal without the additional jets lies below the
continuum background from $q\bar q,gg\to \gamma\gamma$.

We list in Table 4 the number of signal and background events for $WH/ZH$
production for 100
$pb^{-1}$ of integrated luminosity. Table 5 gives the number of signal
and background events for the weak-vector-boson-fusion process, also for
100 $pb^{-1}$. It should be relatively
straightforward to search for a bosonic Higgs decaying to two photons up
to the point where the two-photon branching ratio falls off, roughly
$m_H=100$ GeV\null.  If such a particle were discovered, it would have
dramatic consequences for our understanding of the source of electroweak
symmetry breaking.

\begin{table}[h]
\caption[fake]{Number of signal and background events, per 100 $pb^{-1}$,
for production of a bosonic Higgs in association with a weak vector boson,
followed by $H\to \gamma\gamma$ and $W\to \ell\bar \nu$,
$Z\to \ell\bar \ell,\nu\bar \nu$, or $W,Z\to jj$.}
\bigskip
\begin{center}
\begin{tabular}{cccc}
$m_H$ (GeV) & $WH/ZH$ (leptonic) & $WH+ZH$ (jets) & $jj\gamma\gamma$ \\
60 & 14/11 & 73 & 5 \\
80 & 5.3/5 & 29 & 2.7 \\
100 & 0.8/0.7 & 4.4 & 1.4 \\
\end{tabular}
\end{center}
\end{table}

\begin{table}[hbt]
\caption[fake]{Number of signal and background events, per 100 $pb^{-1}$,
for production of a bosonic Higgs via weak-vector-boson fusion,
followed by $H\to \gamma\gamma$.}
\bigskip
\begin{center}
\begin{tabular}{cccc}
$m_H$ (GeV) & $H+$ 2 jets & $jj\gamma\gamma$ \\
60 & 8 & 4.5 \\
80 & 4.7 & 2.7 \\
100 & 1 & 1.6 \\
\end{tabular}
\end{center}
\end{table}

With 1000 $pb^{-1}$, it may be possible to detect a bosonic Higgs of $m_H>100$
GeV decaying to $WW^*$ if it is produced in association with a $W$ boson, and
the like-sign $W$'s decay leptonically, leading to an isolated
like-sign dilepton plus missing $p_T$
signal.  There is no irreducible background to this signal.  However, the
Higgs-boson mass cannot be reconstructed, due to the loss of the neutrinos.

One might also consider the possibility that the Higgs boson associated with
the generation of the weak-vector-boson masses is also associated with the
top quark, but not with any other fermion (``semi-bosonic'').
This is suggested by the fact
that the top-quark mass is thought to be of the order of the $W$ and $Z$
masses,
and is much heavier than the other known fermions.  Such a Higgs boson might
arise in a top-quark condensate model. The dominant decay of such
a Higgs boson is to two gluons via a top-quark loop for
$m_H=60$--80 GeV, but it still
has a significant branching ratio to two photons.  It will be copiously
produced via gluon fusion.  Since the production cross section is
proportional to the $H\to gg$ partial width, the cross section for
$gg\to H\to \gamma\gamma$ is proportional to $\Gamma(H\to gg)\times
BR(H\to \gamma\gamma) \approx \Gamma (H\to \gamma\gamma)$, i.e., it is
independent of $\Gamma(H\to gg)$, and depends only on $\Gamma(H\to \gamma
\gamma)$.  The cross section for $gg \to H\to \gamma\gamma$ is presented in
Fig.~8, with cuts on the photons of $|\eta_\gamma|<1$ and $p_{T\gamma}>10$ GeV.
The rapidity is restricted to less than unity to suppress the background
from $q\bar q \to \gamma\gamma$, which is peaked in the forward-backward
direction.  This background, combined with $gg \to \gamma\gamma$,
integrated over an invariant-mass bin of $\pm 2\Delta M_{\gamma\gamma}$
centered on $m_H$,
with $\Delta M_{\gamma\gamma}$ as given previously,
is also presented in Fig.~8; it is much larger than the signal.\footnote{We
have not included the QCD correction to $q\bar q \to \gamma\gamma$,
which is given in Ref.~\cite{GAMGAM}.}  More than
1000 $pb^{-1}$ of integrated luminosity would be needed to establish a signal.

\section{Conclusions}

We have studied the production and detection of the standard Higgs boson
at the Fermilab Tevatron.  With 1000 $pb^{-1}$ of integrated luminosity, it
may be possible to observe the Higgs boson produced in association with
a $W$ or $Z$, followed by $H\to b\bar b$ and $W\to \ell\bar \nu$,
$Z\to \ell\bar \ell,\nu\bar \nu$, for $m_H=60$--80 GeV.  Excellent
light-quark-jet/$b$-jet discrimination, and some $c$-jet/$b$-jet
discrimination,
will be required.  Higher Higgs-boson masses will require increased
integrated luminosity for discovery.  Although the search for the Higgs
boson at the Tevatron is tantalizing, our study has further strengthened
our conviction that a higher-energy collider, such as the LHC or SSC,
is needed to explore the electroweak-symmetry-breaking mechanism.

We have also studied the possibility of detecting a non-standard Higgs
boson with suppressed couplings to fermions, dubbed the bosonic Higgs,
via its enhanced two-photon decay mode.  With just 100 $pb^{-1}$ of
integrated luminosity, it will be possible to detect such a particle,
produced in association with a $W$ or $Z$, with the weak vector bosons
decaying leptonically or hadronically, up to $m_H=100$ GeV.  There may
also be an observable signal from the weak-vector-boson-fusion process.
The discovery of such a particle would have an enormous impact on
our understanding of the electroweak-symmetry-breaking mechanism.

\section{Acknowledgements}

We are grateful for conversations with B.~Blair, E.~Boos, S.~Dawson,
E.~Eichten, K.~Ellis,
S.~Errede, G.~Forden, S.~Geer, H.~Gordon, J.~Gunion, T.~Han, R.~Kauffman,
S.~Kuhlmann, T.~LeCompte,
R.~Lipton, T.~Liss, F.~Paige, S.~Parke, S.~Protopopescu, C.~Quigg, A.~White,
and J.~Yoh.  This work was supported under contract no. DE-AC02-76CH00016
with the U.S. Department of Energy.
S.W. thanks the Aspen Center for Physics where part of this work
was performed.  S.W. was partially supported by the Texas National Research
Laboratory Commission.

\clearpage

\section{Figure Captions}

\vrule height0pt
\vspace{-22pt}

\bigskip

\indent Fig.~1 - Cross sections for various Higgs-boson production processes
at the
Tevatron ($\sqrt s = 2$ TeV) versus the Higgs-boson mass.  The HMRSB parton
distribution functions \cite{HMRS} are used for all calculations, and
$m_t=150$ GeV is assumed.

\bigskip

Fig.~2 - Branching ratios of the standard Higgs boson into $b\bar b$, $c\bar
c$,
$\tau^+\tau^-$, $WW^{(*)}$,
and $ZZ^{(*)}$, versus the Higgs-boson mass.

\bigskip

Fig.~3 - Cross sections and backgrounds for a.) $WH$ and b.) $ZH$ production,
followed by $H\to b\bar b$ and $W\to \ell\bar \nu$,
$Z\to \ell\bar \ell,\nu\bar\nu$, versus the Higgs-boson mass.  The cuts which
have been made to simulate the acceptance of the detector are listed in
Table 1. The backgrounds are from $Wb\bar b$ and $Zb\bar b$, $WZ$ and $ZZ$
followed by $Z\to b\bar b$, and $t\bar t \to W^+W^-b\bar b$
with one $W$ missed (for $WH$) or with both $W$'s missed
(for $ZH$, $Z\to \nu\bar \nu$).

\bigskip

Fig.~4 a.) Two-photon decay of a bosonic Higgs via a $W$-boson loop;
b.) Decay of a bosonic Higgs to $b\bar b$ via virtual $W$ and $Z$ bosons.
The latter decay mode is suppressed relative to the former by $m_b^2/M_W^2$.

\bigskip

Fig.~5 - Branching ratios of a bosonic Higgs to $\gamma\gamma$, $WW^{(*)}$,
and $ZZ^{(*)}$, versus the Higgs-boson mass.

\bigskip

Fig.~6 - Cross sections and backgrounds for the bosonic Higgs, produced in
association with a $W$ or $Z$, followed by $H\to \gamma\gamma$ and
(a) $W\to \ell\bar \nu$, $Z\to \ell\bar \ell,\nu\bar \nu$, and
(b) $W,Z\to jj$, versus the Higgs-boson mass.  The cuts which
have been made to simulate the acceptance of the detector are listed in
Table 3.  The signal with the leptonic decay of the weak vector bosons has
no background.  The background to the hadronic decay is from a mixed
QED/QCD process leading to a $jj\gamma\gamma$ final state.

\bigskip

Fig.~7 - Cross section and background for the production of the bosonic Higgs
via weak-vector-boson fusion, followed by $H\to \gamma\gamma$, versus
the Higgs-boson mass.  The cuts which
have been made to simulate the acceptance of the detector are listed in
Table 3.  The two jets in the final state, left over from the radiation of the
weak vector bosons, have an invariant mass greater than 100 GeV.
The background is from a mixed
QED/QCD process leading to a $jj\gamma\gamma$ final state.

\bigskip

Fig.~8 - Cross section and background for the production of the
top-quark-condensate Higgs boson, followed by the decay $H\to \gamma\gamma$,
versus the Higgs-boson mass.  The background is from $q\bar q, gg \to
\gamma\gamma$.

\vfill

\end{document}